\documentstyle[preprint,tighten,aps]{revtex}
\begin{document}
\draft
\title{Spin-gap formation in cuprates: gauge theory}
\author{Menke U. Ubbens and Patrick A. Lee}
\address{
        Dept. of Physics, Massachusetts Institute of Technology,
        Cambridge, MA 02139
}
\date{\today}
\maketitle
\begin{abstract}
We analyze the phase diagram of single and bi-layer cuprates
using the gauge-field description of the $t$-$J$ model.
For $T>T_{\text{BE}}$ the in-plane fermion-pairing order parameter
$\Delta_{\scriptscriptstyle\parallel}$
is eliminated by gauge field fluctuations, leading us to
predict the absence of a spin-gap phase in single-layer cuprates.
For bi-layer cuprates the inter-layer order parameter
$\Delta_\perp$ is enhanced by spin correlations,
and is less affected by gauge fluctuations.
We believe that the spin gap in bi-layer materials
is due to inter-layer fermion pairing, and that for these underdoped
samples the superconducting gap may be nodeless.
\end{abstract}

It is widely accepted that the high-$T_c$ copper-oxide compounds
are Mott-Hubbard insulators which become metallic and
superconducting upon hole doping.
Thus a useful starting point for describing the copper-oxides
is the two-dimensional Hubbard model,
and many workers have studied the strong-coupling limit
which leads to the $t$-$J$ model.
A common technique to incorporate the constraint of no double
occupancy is to decompose the electron operator $c_{i\sigma}^\dagger$
as the product of a fermion operator $f_{i\sigma}^\dagger$
and a boson operator $b_i$, so that the constraint becomes
$\sum_\sigma f_{i\sigma}^\dagger f_{i\sigma} +b_i^\dagger b_i =1$.
In this decomposition the fermions carry the spin, given by
${\bf S}_i= f_{i\alpha}^\dagger \bbox{\sigma}_{\alpha\beta} f_{i\beta}$,
and the bosons represent the empty sites.
The exchange term can be written as
\begin{eqnarray}
{\bf S}_i\cdot{\bf S}_j &=& -\mbox{$\frac12$}(f_{i\sigma}^\dagger f_{j\sigma})
        (f_{j\sigma}^\dagger f_{i\sigma})
        -\mbox{$\frac14$}n_in_j +\mbox{$\frac12$}n_i
\label{phchan}\\
&=& -\mbox{$\frac12$}(f_{i\uparrow}^\dagger f_{j\downarrow}^\dagger
	-f_{i\downarrow}^\dagger f_{j\uparrow}^\dagger)
        (f_{j\downarrow} f_{i\uparrow}-f_{j\uparrow} f_{i\downarrow}),
\label{ppchan}
\end{eqnarray}
which suggests two forms of mean-field decouplings \cite{AnB}:
$\xi_{ij}= \sum_\sigma\langle f_{i\sigma}^\dagger f_{j\sigma} \rangle$
and $\Delta_{ij}= \langle f_{i\uparrow}f_{i\downarrow}
	-f_{i\downarrow}f_{i\uparrow} \rangle$.
Let us briefly discuss the mean-field phase diagram, which is quite simple
for intermediate doping $x\agt 0.04$ \cite{Liu}.
It shows the onset of a nonzero $\xi_{ij}$ below a temperature
$T\sim 0.2 J$, and the onset of $d$-wave fermion pairing
(i.e., $\Delta_{i,i+\hat{x}}= -\Delta_{i,i+\hat{y}}$)
at a lower temperature shown by the dashed line in Fig.\ \ref{phdiag.fig}.
The physical interpretation of this state is the formation
of a singlet ground state (the RVB state \cite{AnB}), with a gap structure
in the spin-excitation spectrum.
In addition, the bosons are expected to effectively Bose condense,
i.e.\ the diamagnetic susceptibility $\chi_B$ will diverge
exponentially below a characteristic temperature, which in the
mean-field approximation is given by
$T_{\text{BE}}^{(0)}= 4\pi t\xi x$.
In the region bounded by $T_{\text{BE}}^{(0)}$ and the onset of
fermion pairing, the pairing amplitude
$\langle c_{i\uparrow} c_{j\downarrow}\rangle$
of the physical electrons is nonzero, and a $d$-wave superconductor
is predicted.
However, above $T_{\text{BE}}^{(0)}$ the holes are incoherent
while the gap structure in the spin-excitation spectrum remains,
and this region of the phase diagram has been called the
{\it spin-gap} phase.

Experimentally there are clear signatures of spin-gap formation
below $T\sim 150 K$ in YBa$_2$Cu$_3$O$_{6+x}$ with $x=0.6$
and in YBa$_2$Cu$_4$O$_8$.
The signature is clearest in the copper and oxygen nuclear
spin-relaxation rates $^{63}T_1^{-1}$ and $^{17}T_1^{-1}$,
but it is also apparent in the Knight-shift data and in $T_2$
measurements \cite{MMP}. Since the spin gap appears only in underdoped
samples, one is naturally led to an explanation in terms of
the fermion pairing discussed above.
However, recently Millis and Monien re-examined the
experimental evidence \cite{MiM}, and concluded that the spin-gap phase is
absent in single-layer material such as La$_{2-x}$Sr$_x$CuO$_4$.
They suggested that the inter-layer exchange coupling
$J_\perp$ may be responsible for the spin-gap phase.
However in the models they constructed, an unreasonably large
exchange $J_\perp$, comparable to $J$, is required to explain the
experimental data.

On the theoretical side, there has been much effort on improving
the mean-field theory by including fluctuation corrections.
It has been recognized that the phase fluctuations of the field
$\xi_{ij}$ are the most important, leading to a
gauge theory in which the fermions and the bosons are coupled
to a U(1) gauge field ${\bf a}_{ij}$ \cite{Bas,Iof}.
Integration over the gauge field enforces the condition that a
flow of fermions is accompanied by a backflow of bosons.
Initially the gauge field has no dynamics,
which it only acquires upon integrating out the fermions and bosons.
We focus on the transverse gauge field because, unlike longitudinal
fluctuations, transverse currents are not screened, giving rise
to unusual infra-red singularities \cite{NL3}.
The transverse gauge-field propagator is given by
$D({\bf q},\omega)= \Pi({\bf q},\omega)^{-1}$, where
$\Pi= \Pi_F+\Pi_B$,
$\Pi_F({\bf q},i\omega_n)= \chi_F q^2 +|\omega_n| k_F/q$,
and $\Pi_B({\bf q},i\omega_n)= \chi_B q^2$.
Here $\chi_F$ and $\chi_B$ are the fermion and boson diamagnetic
susceptibility, and for free bosons
$\chi_B^0= (\exp(T_{\text{BE}}^{(0)}/T)-1)/24\pi m_B$,
while the $|\omega_n|k_F/q$ term represents Landau damping.
Recently we found that gauge fluctuations modify the
mean-field phase diagram in a significant way \cite{MU4,MU3}.
The Bose-condensation temperature is greatly
suppressed, and the spin-gap phase is eliminated entirely.
The former effect is due to thermal fluctuations of the gauge field,
given by
$\langle{\bf h}({\bf x})\cdot{\bf h}(0)\rangle \simeq (T/\chi)\delta({\bf x})$
where ${\bf h}= \nabla\times{\bf a}$ and $\chi= \chi_F+\chi_B$.
Note that at sufficiently low temperatures the RMS flux per
plaquette is small. Nevertheless, since the deBroglie wavelength
$\lambda_T$ of the bosons covers many lattice spacings, the effect
on the random gauge flux is strong.
Indeed we find that the dimensionless coupling constant is
$g= (T/\chi)\lambda_T^2 \sim (m_B\chi)^{-1}$.
At temperatures above Bose condensation we may ignore $\chi_B^0$ compared to
$\chi_F^0$ and we find that $g=24\pi t/J$,
i.e.\ the system is in the strong-coupling
limit even for modest $t/J$.
In this limit the Feynman paths of the bosons are almost self-retracing.
By taking only static gauge-field configurations and taking a
quenched average, we find that $\chi_B$ grows inversely with $T$,
but is suppressed compared to the high-temperature limit
of $\chi_B^0$ by a factor $(0.2 g)^{-1}$ \cite{MU4}.
At sufficiently low temperatures, $\chi_B$ becomes much larger
than $\chi_F^0$ and can
therefore no longer be ignored in the estimate of $g$.
We crossover to the weak-coupling limit and $\chi_B$ self-consistently
diverges exponentially below a temperature $T_{\text{BE}}$
which we estimate to be $T_{\text{BE}}^{(0)}/12$.
This suppression of the effective Bose condensation eliminates
one of the most serious drawbacks of the mean-field phase diagram,
where $T_{\text{BE}}^{(0)}$ yields a temperature scale which
is much too large.

We also examined the effect of gauge fluctuations on the fermion pairing.
Here we find that it is the quantum fluctuations which are important,
and the effect is strong enough to de-stabilize the mean-field
transition until the temperature falls below $T_{\text{BE}}$ \cite{MU3}.
We computed the gauge field contribution to the free energy
\begin{equation}
F_{\text{gauge}}=
\sum_{\bf q} \int_0^\infty \frac{d\omega} {2\pi}
        [2 n_B(\omega)+1] \arctan\left(
        \frac{\mbox{Im}\> \Pi({\bf q},\omega+i\delta)}
        {\mbox{Re}\> \Pi({\bf q},\omega+i\delta)} \right).
\label{Fgauge}
\end{equation}
For $T>T_{\text{BE}}$, $\Pi$ may be replaced by $\Pi_F$, and
Eq.\ (\ref{Fgauge}) yields a large negative contribution to the
free energy which is cutoff by $T$, leading to a $T^{5/3}$ correction
term, and a specific heat of $T^{2/3}$ \cite{NL3}.
Now we can ask what happens if the fermions form a $d$-wave pairing state
with amplitude $\Delta$. We can compute $F_{\text{gauge}}(\Delta)$
by replacing $\Pi$ in Eq.\ (\ref{Fgauge}) by the corresponding
polarization function $\Pi({\bf q},i\omega_n,\Delta)$ in the
pairing state. Since a gap appears in $\mbox{Im}\>\Pi({\bf q},\omega)$,
the $\omega$ integral is cutoff by $\Delta$ instead of $T$,
and we obtain $F_{\text{gauge}}\sim \Delta^{5/3}$.
This is to be added to the mean-field gain in free energy
$F_{\text{MF}} \sim -\Delta^2$, which is overwhelmed by
$F_{\text{gauge}}$ for small $\Delta$.
However, a first-order transition
is in principle still possible. We have carried out a detailed
numerical computation of $F_{\text{gauge}}(\Delta)$ \cite{MU3},
and found that the pairing state is unstable for $T>T_{\text{BE}}$.
Below $T_{\text{BE}}$, $\chi_B(T)$ grows exponentially and
consequently $\mbox{Re}\>\Pi_B$ dominates over $\mbox{Re}\>\Pi_F$.
Thus $F_{\text{gauge}}(\Delta)$ becomes less important and the cost
of opening up a gap is correspondingly weaker.
Our calculations show that a first-order transition to a $d$-wave
pairing state occurs below $T_{\text{BE}}$.
Since the bosons are coherent, this is a direct transition to a
$d$-wave superconductor. Our best estimate of the transition
temperature is shown in Fig.\ \ref{phdiag.fig}.
We expect the first-order nature of the transition to be smoothened
out once phase fluctuations of the superconducting order parameter
are taken into account.
An important consequence of this analysis is that the spin-gap phase
is completely eliminated by gauge fluctuations.
This agrees with the absence of a spin-gap phase in single-layer
materials such as La$_{2-x}$Sr$_x$CuO$_4$.
%

Now that we have de-stabilized the spin-gap phase in a single layer,
we have to appeal to inter-layer coupling to explain the
appearance of spin gaps \cite{MiM}.
In a bi-layer material it is natural to look for an order parameter \cite{AI2}
\begin{equation}
\Delta_\perp({\bf r}_{ij})=
\langle f_{i\uparrow}^{(1)}f_{j\downarrow}^{(2)}
-f_{i\downarrow}^{(1)}f_{j\uparrow}^{(2)} \rangle,
\label{D12}
\end{equation}
by decoupling the inter-layer exchange term
$H_\perp= J_\perp\sum_i {\bf S}_i^{(1)} {\bf S}_i^{(2)}$.
At present we only have a lower bound $J_\perp\agt \mbox{8 meV}$
from the absence of an optical mode in the neutron scattering
from undoped YBCO \cite{Tra92}, but we expect $J_\perp$ to be much smaller
than $J$. A conventional pairing theory would then predict a gap
which is exponentionally small in $J/J_\perp$.
To overcome this we note that the underdoped cuprates have a
reasonable long antiferromagnetic correlation length in the plane.
In a bi-layer it is then favorable to form singlets not only between
spins which are directly on top of each other, but also between
other pairs on the same sublattice.
Thus we expect the inter-layer pairing
gap to be enhanced by in-plane correlations.
To analyze this we employ the random-phase approximation (RPA),
which has been quite succesful in reproducing the experimentally
observed spin correlations \cite{Tan}.
The RPA approach corresponds to a decoupling of the
exchange term ${\bf S}_i\cdot {\bf S}_j$ in a third channel which
differs from the decouplings in Eqs.\ (\ref{phchan}) and (\ref{ppchan}).
In order to avoid double counting we separate the Hamiltonian into
two equal parts, decouple the first part using Eq.\ (\ref{phchan})
and apply RPA on the second part.
Our results should not be too sensitive to this admittedly
{\it ad hoc} procedure. Within this scheme we define
$J_\perp^{\text{eff}}({\bf q})$ as the effective inter-layer
exchange due to the summation of a series of RPA bubbles
shown in Fig.\ \ref{Jk.fig}.
This analysis will be discussed in more detail in Ref.\ \cite{MU5}.
The result is
\begin{equation}
J_\perp^{\text{eff}}({\bf q})= \frac{J_\perp}
{(1+\frac14 \chi^0J_{\scriptscriptstyle\parallel}({\bf q}))^2
-(\frac14\chi^0J_\perp)^2},
\label{Jperp}
\end{equation}
where $\chi^0({\bf q})$ is the free-fermion bubble for the
two-dimensional band structure.
The denominator vanishes at the onset of the N\'{e}el ordering,
which occurs at $x_c\simeq 0.08$ within our approximation.
The numerical solution of $J_\perp^{\text{eff}}({\bf q})$
for $x\agt x_c$ is shown in Fig.\ \ref{Jk.fig},
and we see that it is significantly
enhanced near the incommensurate nesting vector
${\bf Q}_{\text{AF}}\simeq (\pi,\pi\pm 2x)$.
We find that for $x=0.09$ the peak at ${\bf Q}_{\text{AF}}$ leads
to an antiferromagnetic correlation length of approximately
3 lattice spacings.
We next take the Fourier transform of $J_\perp^{\text{eff}}({\bf q})$
to obtain $J_\perp^{\text{eff}}({\bf r}_{ij})$ and consider
the effective inter-layer exchange
$J_\perp^{\text{eff}}({\bf r}_{ij}) {\bf S}_i^{(1)}\cdot {\bf S}_j^{(2)}$,
which is decoupled using Eqs.\ (\ref{ppchan}) and (\ref{D12}).
The gap equation for the order parameters $\Delta_\perp({\bf r}_{ij})$
is solved numerically in real space.
We find that the pairing amplitude is especially large on the
diagonal such that
$\Delta_\perp({\bf r}_{ii})\simeq |\Delta_i| (-1)^i$
decays slowly with $i$.
Transforming $\Delta({\bf r}_{ij})$ back to $k$ space, we find a
quasi-particle dispersion
$E({\bf k})= [\epsilon({\bf k})^2 +\Delta_\perp({\bf k})]^{1/2}$,
where $\Delta_\perp({\bf k})$ is shown in Fig.\ \ref{Dk2D.fig}.
Notice that $\Delta_\perp({\bf k})$ exhibits some anisotropy
but does not change sign.
Thus the inter-layer pairing is $s$-wave in nature,
with a full gap in the excitation spectrum.

It is important to note that it is crucial in this
approach that the in-plane $d$-wave pairing is suppressed by
gauge fluctuations.
One might ask whether gauge fluctuations will destroy the inter-plane
pairing as well. We believe that this does not happen for the
following reason. There are two gauge fields modes $a_1$ and $a_2$
in the two layers, which are split into
$a_\pm\equiv (a_1\pm a_2)/\sqrt{2}$ at the onset of inter-layer
pairing. The gauge propagators are given by $\Pi_\pm^{-1}$ where
\begin{eqnarray}
\Pi_\pm({\bf q},i\nu_n)&=& C+ 2T\sum_{i\omega_n}\int\frac{d^2k}{(2\pi)^2}
\left(\hat{\bf q}\times\frac{\partial\epsilon}{\partial{\bf k}}\right)
\left(\hat{\bf q}\times\frac{\partial\epsilon'}{\partial{\bf k}}\right)
\nonumber\\
&&\times\frac{\epsilon\epsilon' -\omega_n\omega_n' \pm\Delta\Delta'}
{(\omega_n^2 +E^2)(\omega_n'^2 +E'^2)},
\label{Pipm}
\end{eqnarray}
where $i\omega_n'=i\omega_n-i\nu_n$;
$\epsilon,\epsilon'=\epsilon({\bf k}\pm {\bf q}/2)$;
$\Delta,\Delta'=\Delta_\perp({\bf k}\pm {\bf q}/2)$;
and $E=\sqrt{\epsilon^2 +\Delta^2}$.
For $\Delta=0$, the constant $C$ exactly cancels the second term
for ${\bf q}\to 0$ and $\nu_n=0$, so that the gauge field is
massless in the normal state.
In the pairing state a gap opens up in $\Pi_+(0,0)\sim \Delta_\perp^2$,
but $\Pi_-$ remains massless.
Furthermore, when we substitute $\Pi_\pm$ into Eq.\ (\ref{Fgauge}),
we find that while the $\Pi_+$ propagator introduces an energy cost
and is pair-breaking as before, the $\Pi_-$ propagator
is actually {\it pair-enhancing}. This is because even though an
energy gap is introduced into $\mbox{Im }\Pi_-$, the coherence
factor in Eq.\ (\ref{Pipm}) is such that
$-(\mbox{Im}\>\Pi_-/\mbox{Re}\>\Pi_-)$ is {\it enhanced}
for $\omega> 2\Delta_\perp$ compared to its normal state value,
and this overwhelms the loss of the contribution from $\omega<2\Delta_\perp$.
This pair-enhancing nature of the $a_-$ mode can also be
understood in another way \cite{BoS}.
The fermions on the two planes couple to the $a_-$ mode with
opposite charge, so that the exchange of an $a_-$ mode leads
to an attraction, analogous to what happens in the
$t$-$t'$-$J$ model \cite{WW}.
In our case we expect that the effects of the $a_+$ and $a_-$
gauge fluctuations largely cancel each other,
leaving the mean-field transition intact.

Since we are unable to compute $\Delta_\perp$ quantitatively,
in the remainder of the paper we will use $\Delta_\perp$ as a parameter
to compute various physical quantities, which we compare with
experiments. We calculated the nuclear-relaxation rate $1/T_1$
for the copper and the oxygen sites, and we immediately
encountered the problem that for $s$-wave pairing the
Hebel-Slichter peak appears below $T_c$.
We circumvented this by arguing that since the order parameter is not
invariant under the local gauge transformation
$f_{i\sigma}^{(1,2)}\to e^{i\theta_i} f_{i\sigma}^{(1,2)}$
the transition should be broadened into a crossover behavior,
and a BCS-type gap should be replaced by a pseudo gap
as shown in the inset in Fig.\ \ref{T1T.fig}.
With this and a modest broadening $\Gamma= 0.1 T$, we eliminated
much of the Hebel-Slichter peak, and the results are shown
in Fig.\ \ref{T1T.fig}.
Notice that above the crossover transition, $(^{63}T_1T)^{-1}$
increases upon lowering $T$, which is due to antiferromagnetic correlations.
We do not see this increase in $(^{63}T_1T)^{-1}$, because the
form factor $^{17}F({\bf q})$ for the oxygen site vanishes
at the antiferromagnetic nesting vector ${\bf Q}_{\text{AF}}$ \cite{MMP}.
We also computed $1/T_2$ and the uniform susceptibility
$\chi({\bf q}=0)$, shown in the second panel of Fig.\ \ref{T1T.fig}.
The results compare reasonably well with experiments provided
we choose $\Delta_\perp\simeq 150 K\simeq 0.1 J$.
We believe that this is not an unreasonable value if $J_\perp\simeq 0.2J$.

So far we have ignored any inter-layer hopping of the form
$t_\perp c_{i\sigma}^{(1)\dagger} c_{i\sigma}^{(2)}$,
which is reasonable provided $xt_\perp< J_\perp$.
If this is violated we expect inter-layer pairing to be
suppressed, but we have not studied this quantitatively.
Not enough is known about $t_\perp$ and $J_\perp$, but our guess
is that $xt_\perp$ and $J_\perp$ are comparable.
However, even a small $t_\perp$ will lead to coherence between
bosons on the two planes immediately below $T_{\text{BE}}$, so that
the fermion pairing becomes genuine superconducting pairing
between electrons on the two layers.
At low temperatures the in-plane $s$-wave and the inter-plane
$d$-wave pairing will co-exist, giving rise to a quasi-particle
dispersion $E({\bf k})= (\epsilon({\bf k})^2 +\Delta_\pm({\bf k})^2)^{1/2}$,
where $\Delta_\pm({\bf k})= \Delta_\perp({\bf k})\pm
\Delta_{\scriptscriptstyle\parallel}({\bf k})$.
If $\Delta_\perp$ is indeed as large as 150 K, as the experiments seem
to indicate, comparison with Fig.\ \ref{phdiag.fig} indicates
that for underdoped materials it is likely that
$\Delta_\perp> |\Delta_{\scriptscriptstyle\parallel}|$
for all ${\bf k}$,
so that the superconducting gap is nodeless.
As doping is increased, $\Delta_\perp$ decreases rapidly with
the loss of antiferromagnetic correlations, which was essential
for the enhancement of $J_\perp$,
and we crossover to a superconducting state with nodes.
As long as $\Delta_\perp$ remains finite we expect that there
are eight nodes along the Fermi surface instead of the four nodes
for a conventional $d$-wave superconductor.
This new phase diagram is shown schematically in the insert
in Fig.\ \ref{phdiag.fig}.
Finally we note that Chakravarty {\it et al.} \cite{Cha}
have shown that inter-layer pair tunneling leads to an
anisotropic $s$-wave gap function.
We emphasize that in our theory the order-parameter symmetry
is always $d$-wave in the plane, even in the nodeless state.
Furthermore, to the extent that our nodeless pairing state is
always preceded by a spin-gap phase, we believe that
the origin of superconductivity in our treatment is closer to
the original RVB picture of pre-formed spin-singlet pairs \cite{AnB}
than the more recent inter-layer pair-tunneling model.

We acknowledge helpful conversations with A.J. Millis.
This work was supported by the NSF through the Material Research
Laboratory under Grant No.\ DMR-90-22933.

\newpage
\begin{figure}
\caption{
The phase diagram for single-layer cuprates.
The dashed line is the mean-field pairing transition.
The solid line takes gauge fluctuations into account, so that
fermion pairing only survives for $T<T_{\protect\text{BE}}$,
resulting in a $d$-wave superconductor.
The inset is a schematic phase diagram for bi-layer materials.
The spin-gap phase is due to inter-layer pairing.
The superconducting state has $d$ symmetry in the plane with
nodes in the gap function, but there may be a region
indicated by the shaded area where the gap is nodeless.
\label{phdiag.fig}}
\end{figure}

\begin{figure}
\caption{
The effective interlayer coupling $J_\perp^{\protect\text{eff}}({\bf q})$
for various values of the doping $x$, using $J_\perp= 0.2 J$.
This is obtained by summing over RPA bubbles (inset).
For $x \to 0.08$ there are strong incommensurate peaks at the nesting vector
${\bf Q}_{\protect\text{AF}}$.
\label{Jk.fig}}
\end{figure}

\begin{figure}
\caption{
The gap $\Delta_\perp({\bf k})$ for $x=0.085$ and $J_\perp^0= 0.2 J$.
The gap has an extended $s$-wave symmetry, and is enhanced close
to the Fermi surface (dotted line), especially at the corners.
\label{Dk2D.fig}}
\end{figure}

\begin{figure}
\caption{
The first panel shows the nuclear-relaxation rates $(T_1T)^{-1}$ on the
$^{63}$Cu and the $^{17}$O sites.
The rise is $(^{63}T_1T)^{-1}$ is due to antiferromagnetic correlations,
which is absent in $(^{17}T_1T)^{-1}$.
The second panel shows $1/T_2$ and $\chi(0,0)$.
The opening of a pseudo-gap (see inset) causes a rapid decrease in
$(T_1T)^{-1}$, $1/T_2$, and $\chi(0,0)$ for $T\protect\alt 0.1 J$.
\label{T1T.fig}}
\end{figure}

\end{document}